\documentstyle[twoside,fleqn,espcrc2,epsf]{article}

\newcommand{\beq}{\begin{equation}}
\newcommand{\eeq}{\end{equation}}
\newcommand{\bea}{\begin{eqnarray}}
\newcommand{\eea}{\end{eqnarray}}
\newcommand{\bi}{\bibitem}
\newcommand{\ol}[1]{\overline{#1}}

\newcommand{\NP}{Nucl.\ Phys.\ }
\newcommand{\PL}{Phys.\ Lett.\ }
\newcommand{\PR}{Phys.\ Rev.\ }

\title{
\vspace{-3.5cm} 
\begin{flushright}
{\normalsize\sc RIKEN BNL Research Center preprint}\\
\end{flushright}
\vspace*{2.3cm}
Domain wall fermions and the strange quark mass
\thanks{Talk presented at QCD 99, Montpellier, France}
}

\author{Matthew Wingate\address{RIKEN BNL Research Center,
        Brookhaven National Laboratory, Upton, NY 11973, USA}
}

\begin{document}

\begin{abstract}
The strange quark mass has been computed using a lattice action
which possesses continuum--like chiral symmetry to good precision, 
namely the domain wall fermion action.  This talk surveys this 
action and the recent calculation of $m_s$ by the RIKEN/BNL/CU 
collaboration.  This result is put into context by briefly summarizing
other recent lattice studies.
\end{abstract}

\maketitle

\section{INTRODUCTION}

The strange quark mass is a fundamental parameter of the
Standard Model, one whose value is poorly known.
The two methods which have been used extensively to determine
$m_s$ are QCD sum rules and lattice QCD simulations.
Before one can compare results from these different
methods fairly, a basic understanding of each method's systematic
uncertainties and approximations is necessary.  This
talk is focused on recent lattice calculations of $m_s$ emphasizing
such uncertainties.  The calculation
of $m_s$ using domain wall fermions by the RIKEN/BNL/Columbia
(RBC) lattice collaboration~\cite{ref:RBC_COLLAB} is taken as a case
study.  The details of this work
were presented at Lattice 99~\cite{ref:MW_LAT99},
and there was another description of the relevant lattice techniques
here by Becirevic~\cite{ref:DB_QCD99}.

\section{DOMAIN WALL FERMIONS}

Symmetries are of fundamental importance in physics; in the case
of QCD chiral symmetry is responsible for much of light hadron dynamics.
However, chirally symmetric discretizations of the Dirac operator
suffer from a ``doubling'' of the particle spectrum.  For example replacing
$\partial_\mu$ with a nearest--neighbor difference operator leads to
the free quark propagator $S(p) = [i\gamma_\mu\sin(p_\mu a) + m]^{-1}$
which, in the massless limit, has poles at all corners of the Brillouin zone
$p_\mu\in[0,\pi]$.  Wilson added to the lattice action a 
second derivative--like operator which breaks
chiral symmetry and gives the doubler states a mass inversely 
proportional to the lattice spacing.  Another approach, 
by Kogut and Susskind (KS),
is to stagger the spin degrees of freedom and interpret the doublers as
different quark flavors.  Although this method maintains a $U(1)$ remnant of
the continuum chiral symmetry, flavor symmetry is badly broken.
Both actions recover the symmetries of continuum quarks only
in the $a\to 0$ limit.

\begin{figure}
\vspace{1.5in}
\includegraphics{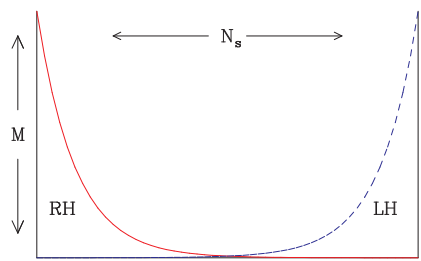}
\caption{Schematic drawing of the light fermion wavefunctions
in the extra dimension.  For appropriate choice of the domain wall
height $M$, a right--handed (RH) mode appears on the left boundary
and a left--handed (LH) mode on the right boundary, $N_s$ sites away.
The mixing between the two modes within the extra dimension decays
exponentially as $N_s$ increases.  }
\label{fig:dwalls}
\end{figure}

An exciting and fruitful suggestion for decoupling the chiral and
continuum limits started with Kaplan~\cite{ref:KAPLAN} who
pointed out to the lattice community that a chiral $2k$--dimensional mode
is bound to a mass defect (or {\it domain wall}) 
in a $2k+1$--dimensional space.
Specifically, in Shamir's boundary fermion 
variant~\cite{ref:SHAMIR,ref:FURSHAM},
there is a massless right--handed (RH) mode stuck to the $4d$
boundary of a semi-infinite $5d$ space, given an appropriate choice
for the $5d$ mass $M$.  When the fifth dimension is made finite,
a left--handed (LH) mode appears on the other boundary (see 
Fig.~\ref{fig:dwalls}).  The wavefunctions
of the light modes decay exponentially within the fifth dimension,
so there is a residual mixing between the RH and LH states which
decreases as one increases $N_s$, the number of sites in the fifth
dimension.  This coupling of chiral modes describes a Dirac fermion
with mass $m_{\rm res}$.  
In order to have control over the mass of the light Dirac fermion,
the RH and LH boundaries are explicitly
coupled with a weight $-am$.  For $am\gg am_{\rm res}$
the mixing within the bulk of the fifth dimension is negligible.

The important feature of the domain wall (DW) fermion action is 
that the continuum limit $a\to 0$
is decoupled from the chiral limit $N_s\to\infty$ at the expense
of an extra dimension of size $N_s$.  Simulations with
DW fermions have shown that for lattice spacings roughly
0.1 fm, $N_s = 10-20$ is sufficient for lattice QCD to exhibit
continuum--like chiral properties, e.g.\ Ward 
identities~\cite{ref:BLUM_SONI} and suppression of wrong--chirality
operator mixings~\cite{ref:CD_LAT99}.

\section{RESULTS OF RBC SIMULATIONS}

There are two ingredients which
contribute to the determination of quark masses from the lattice:
the calculation of hadron masses, decay constants, and/or 
matrix elements; and the calculation of the quark mass 
renormalization constant.  

In Refs.~\cite{ref:MW_LAT99,ref:LW_LAT99} the
pseudoscalar meson, vector meson, and nucleon masses are 
computed as 
functions of the bare quark mass, $am$.  The $am$ corresponding
to the light quark mass is determined by extrapolating
$(aM_{\rm PS}/aM_{\rm V})^2$ to $(M_\pi/M_\rho)^2$, and then
the lattice spacing $a$ can be defined by setting the $\rho$
or $N$ to its physical value.
Finally the bare $am$ corresponding
to the strange quark mass is fixed by interpolating the pseudoscalar
mass to $M_K$ or the vector mass to $M_{K^*}$ or $M_\phi$.

\begin{figure}
\vspace{3.0cm}
\includegraphics{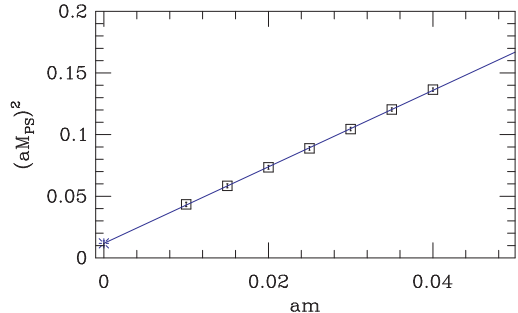}
\caption{Pseudoscalar meson mass squared vs.\ $am$.}
\label{fig:mpi2_m_jk_19jun}
\end{figure}

\begin{figure}
\vspace{3.0cm}
\includegraphics{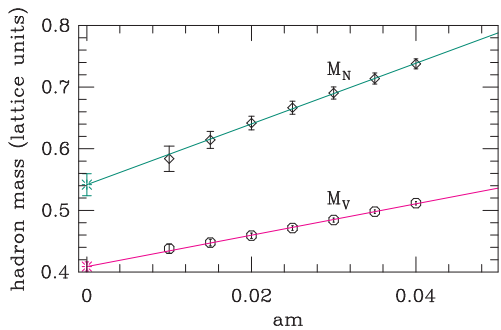}
\caption{Vector meson and nucleon masses vs.\ $am$.}
\label{fig:mhad_m_jk_25jun}
\end{figure}

Figs.~\ref{fig:mpi2_m_jk_19jun} and \ref{fig:mhad_m_jk_25jun}
show these hadron masses as functions of $am$.  Due to a 
combination of finite spatial volume effects, finite $N_s$
effects, and quenching errors, the chiral limit is at
$am= -0.0038(6)$.  Interpolating in the hadron masses as described
above, one finds (with statistical errors only) \cite{ref:MW_LAT99}:

\vspace{0.3cm}
\begin{tabular}{c|cc}
bare & \multicolumn{2}{c}{ {$1/a = 1.91(0.04)$ GeV}} \\
mass & lattice units & MeV \\ \hline
$m_l$ & 0.00166(0.00005) &  {3.17(0.11)} \\ 
$m_s(K)$ & 0.042(0.003) &  {80(6)} \\ 
$m_s(\phi)$ & 0.053(0.004) &  {101(7)} \\ 
\end{tabular}
\vspace{0.3cm}

\noindent
The scale above has been defined by setting the $\rho$ mass to 
its experimental value; however $M_N/M_\rho = 1.37(5)$ compared
to the experimental ratio 1.22, so there is at least a 10\%
uncertainty due to the ambiguity in whether the nucleon or $\rho$
is used to set the scale.  This is common among quenched lattice
simulations with lattice spacings $O(0.1{\rm fm})$.

In Ref.~\cite{ref:MW_LAT99} the renormalization constant is
computed nonperturbatively
using the regularization independent momentum subtraction (RI/MOM)
scheme as advocated by the Rome--Southampton group~\cite{ref:ROME}, 
applied for the first time to DW fermions~\cite{ref:CD_LAT99}.
The result at 2 GeV for $Z^{\ol{\rm MS}}_m
\equiv 1/Z^{\ol{\rm MS}}_S$ is $1.63 \pm 0.07$ (stat.)\ $\pm 0.09$ (sys.)
which should be compared to 1.32, the one--loop perturbative $Z_m$
\cite{ref:BSW,ref:AIKT} for these simulation parameters.
One inherent assumption of this method is that perturbation
theory is reliable at the scale of the inverse lattice spacing,
roughly 2 GeV here.  As Chetyrkin pointed out~\cite{ref:CHETYRKIN}
the convergence of the perturbation series for the matching
between RI/MOM and ${\ol{\rm MS}}$ schemes is slow at 2 GeV.
The convergence would be much improved for $1/a \approx 2.5-3$ GeV.

The result from Ref.~\cite{ref:MW_LAT99} at this lattice spacing 
(corresponding to $6/g_0^2 = 6.0$) is
\beq
m^{\ol{\rm MS}}_s(2 ~{\rm GeV}) = 130\pm 11 \pm 18 ~{\rm MeV},
\eeq
where the first error is the statistical uncertainty 
and the second is the systematic uncertainty.

\section{COMPARISON OF RESULTS}

\begin{figure}
\vspace{2.5in}
\includegraphics{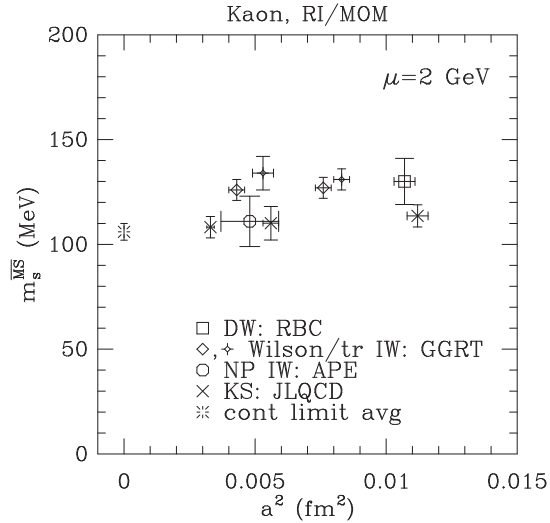}
\caption{Strange quark mass vs.\ lattice spacing.  
Except for $a^2=0$ point, all $m_s$ values computed using the
$\rho$ to set the scale, the $K$ to set $m_s$, and the RI/MOM
scheme.  The $a^2=0$ asterisk is an average
of three recent continuum limit calculations.
See text for further description.}
\label{fig:ms_k_npr_final.av}
\end{figure}

Fig.~\ref{fig:ms_k_npr_final.av} shows several quenched lattice
calculations of $m_s$ plotted vs.\ lattice spacing squared.
Except for the point at $a^2=0$, the quark masses were determined
using the procedure described above and are plotted with 
statistical error bars only.  The square corresponds
to the DW fermion simulation of Ref.~\cite{ref:MW_LAT99}.
The crosses are from KS simulations~\cite{ref:JLQCD_NPR},
the big (small) diamonds are from Wilson (tree--level improved Wilson)
simulations~\cite{ref:GGRT}, and the octagon from a nonperturbatively
improved Wilson action~\cite{ref:BBLLMM}.  Finally, the asterisk
displays the author's average of three recent continuum extrapolations
of $m_s$, as described in the following paragraph.

Recently there have been three quenched lattice calculations of $m_s$ 
which use nonperturbative renormalization and take the continuum limit:
one with KS fermions~\cite{ref:JLQCD_NPR} and two with nonperturbatively
improved Wilson (NP IW) fermions~\cite{ref:ALPHA_UKQCD,ref:QCDSF99},
(see table below).
In order to precisely compare among different lattice simulations which
contain similar, if not identical, systematic effects, it is 
common practice to quote only statistical errors.  However, before 
using the results phenomenologically, one must do one's best to
include these systematic errors.  As these simulations were
performed on large volumes, the source of the largest systematic
error (within the quenched approximation) is the ambiguity
of which physical quantity is used to set the lattice spacing.
Ref.\ \cite{ref:JLQCD_NPR} uses the lattice spacing set by 
the $\rho$ mass, $a(M_\rho)$, while 
Refs.\ \cite{ref:ALPHA_UKQCD,ref:QCDSF99} use $a(f_K)$ (obtaining
identical results with the lattice spacing set by the
hadronic radius parameter $r_0$) and 
 report that using $a(M_N)$ instead would increase $m_s$
by 10 MeV.  It is commonly seen in quenched lattice QCD that
\beq
a(f_K) < a(M_\rho) < a(M_N).
\eeq
Since the latter two references quote a one--sided systematic uncertainty,
the central value is shifted up by half of that uncertainty and the new error
contains the statistical and systematic uncertainties added in quadrature.
A 5\% scale uncertainty is added to the statistical error of
Ref.\ \cite{ref:JLQCD_NPR}.
The weighted average of the three results is $106 \pm 4$ MeV.

\noindent
\vspace{0.2cm}
\begin{tabular}{c|cc|c}
Ref.\ & quoted $m_s$ & $a^{-1}$ syst.\ & adjusted $m_s$ \\ \hline
\cite{ref:JLQCD_NPR} & 106(7) MeV  &  $\pm 5$ MeV & 106(9) MeV \\
\cite{ref:ALPHA_UKQCD} & ~97(4) MeV & +10 MeV & 102(6) MeV  \\
\cite{ref:QCDSF99} & 105(4) MeV  & +10 MeV  & 110(6) MeV \\ \hline
\multicolumn{3}{c}{weighted average} & 106(4) MeV
\end{tabular}
\vspace{0.3cm}

Of course in nature there are sea quarks,
so the quenched approximation could prove unreliable.  Last year the
CP-PACS collaboration reported deviations from experiment 
of the quenched hadron spectrum in the continuum limit at the
level of 10\%~\cite{ref:CPPACS_LAT98}, one might expect the same level of 
disagreement for the strange quark mass.  However, this year the
CP-PACS collaboration has a preliminary result from two--flavor 
dynamical Wilson fermion simulations of 
$m_s^{\ol{\rm MS}}(2~{\rm GeV}) = 84\pm7$ MeV~\cite{ref:CPPACS_LAT99}.
Although the mass renormalization has been computed only in 
(improved) perturbation theory, such a small strange quark mass
is in conflict with lower bounds based on the positivity of 
hadronic spectral functions~\cite{ref:LRT}.

\section{CONCLUSIONS}

The RBC collaboration~\cite{ref:RBC_COLLAB} 
has completed the calculation of the
strange quark mass at a single gauge coupling $6/g_0^2 = 6.0$
using the domain wall fermion action.  The mass renormalization
is determined nonperturbatively through the RI/MOM scheme.
The result, $130\pm11\pm18$ MeV, is consistent others at the
same gauge coupling.  Note this is a result at only one lattice
spacing.  Only in the continuum limit can
a reliable comparison can be made to the NPSW and
KS average of $106 \pm 4$ MeV.
A final answer for $m_s$ awaits several unquenched simulations 
exploring all the different systematic uncertainties already probed
within the quenched approximation.

\section*{ACKNOWLEDGMENTS}

The author thanks the organizer of this conference for a 
stimulating forum for discussion.  Thanks also to RIKEN,
Brookhaven National Lab, and the U.S.\ Department of Energy for 
providing the facilities essential for
the completion of this work.

\end{document}